\documentclass[twocolumn]{aa}
\usepackage{graphicx}
\def\ABDorA{AB\,Dor\,A}
\def\ABDorB{AB\,Dor\,B}
\def\ABDorC{AB\,Dor\,C}
\def\ABDorBb{AB\,Dor\,Bb}
\def\ABDorBa{AB\,Dor\,Ba}
\def\ABDoradus{AB\,Doradus}

\def\Rst137B{Rst\,137\,B}
\def\orb{S}

\begin{document}

\title{On the dynamics of the AB\,Doradus system}

\author{J.C. Guirado\inst{1}, I. Mart\'{\i}-Vidal\inst{1}, J.M. Marcaide\inst{1}, 
	L.M. Close\inst{2}, J.C. Algaba\inst{1}, W. Brandner\inst{3}, 
        J.-F. Lestrade\inst{4}, D.L. Jauncey\inst{5}, D.L. Jones\inst{6}, 
        R.A. Preston\inst{6}, and J.E. Reynolds\inst{5} }

\offprints{J.C. Guirado}

\institute{Departamento de Astronom\'{\i}a y Astrof\'{\i}sica, Universidad de Valencia,
E-46100 Burjassot, Valencia, Spain
\and Steward Observatory, University of Arizona, Tucson, Arizona 85721, USA 
\and Max-Planck Institut f\"ur Astronomie, K\"onigstuhl 17, 69117 Heidelberg, Germany
\and Observatoire de Paris/LERMA, Rue de l'Observatoire 61, F-75014, Paris, France 
\and Australian Telescope National Facility, P.O. Box 76, Epping, NSW 2121, Australia 
\and Jet Propulsion Laboratory, California Institute of Technology, 4800 Oak Grove Drive, 
Pasadena, California 91109, USA 
}
%
		     %

\date{Received July 4, 2005; accepted September 21, 2005 }

\abstract{
We present an astrometric analysis of the binary systems 
AB\,Dor\,A\,/AB\,Dor\,C and \ABDorBa\,/AB\,Dor\,Bb. These two systems of 
well-known late-type stars are gravitationally 
associated and they constitute the quadruple AB\,Doradus system. 
From the astrometric data available at different wavelengths, 
we report:
(i) a determination of the orbit of AB\,Dor\,C, 
the very low mass companion to AB\,Dor\,A, which confirms the mass estimate of 
0.090M$_\odot$ reported in previous works; 
(ii) a measurement of the parallax of AB\,Dor\,Ba, which unambiguously confirms the long-suspected physical association between this star and \ABDorA; 
and (iii)  evidence of orbital motion of \ABDorBa\, around \ABDorA, which 
places an upper bound of 0.4\,M$_\odot$ on the mass of the pair 
\ABDorBa\,/\ABDorBb\, (50\% probability).
Further astrometric monitoring of the system at all possible 
wavelengths would determine with extraordinary precision the dynamical mass 
of its four components.

   }
\titlerunning{On the dynamics of the AB\,Doradus system}
\authorrunning{Guirado et al.}
   \maketitle
%

\section{Introduction}

Astrometry, one of the most classical astronomical disciplines, 
provides unambiguous mass estimates 
of celestial bodies via observations of the orbits 
of binary or multiple systems, as projected on the sky 
(see, e.g., Kovalevsky 1995).
The precise determination of stellar masses is fundamental in 
astronomy, as this parameter is the primary input to 
test stellar evolutionary models that provide 
widely used mass-luminosity relations.  
In particular, the calibration of the mass-luminosity 
relation for the lower end of the main sequence is 
of special interest, since it permits the derivation of the 
physical properties of very-low-mass (VLM) stars and 
substellar objects. 
However, a model-independent measurement of the mass of these 
objects is a most demanding task that requires the 
astrometric follow-up of VLM stars in binary systems 
(e.g. Lane et al. 2001; Bouy et al. 2004; Golimowski et al. 2004; 
Close et al. 2005).
One of the few VLM objects with astrometric detections  
is \ABDorC, the companion to \ABDorA.\\
\ABDorA\, (=HD\,36705) is an active K1 star only 
14.9 pc away from the Sun. Due to its ultrafast rotation
(0.514 days; Innis et al. 1985), \ABDorA\, is a strong emitter at 
all wavelengths, and it has been extensively 
observed from radio to X-rays (Lim et al. 1992; Mewe et al. 1996; 
Vilhu et al. 1998; G\"udel et al. 2001).
\ABDorA\, possesses a low-mass companion, \ABDorC, which induces a reflex motion 
first detected by very-long-baseline-interferometry (VLBI) and the Hipparcos satellite 
(Guirado et al. 1997). Recently, Close et al. (2005) [CLG] obtained a near-infrared 
image of \ABDorC, providing the first dynamical calibration of the mass-luminosity 
relation for low mass, young objects. 
\ABDorA\, has another physical companion, 
AB\,Dor\,B (=Rossiter\,137\,B, =\Rst137B), a dM4e star, which is also a rapid rotator 
with a 0.38 day period and is separated from \ABDorA\, by 9" (Lim 1993). 
Based on their young age (CLG), common proper motions, and common 
radial velocities (Innis, Thompson \& Coates 1986), 
it is believed that both stars may be associated. In turn, CLG 
found \ABDorB\, to be a tight binary (\ABDorB=\ABDorBa\, and \ABDorBb).\\
\ABDorC\, is the first calibration point for evolutionary tracks in the young 
VLM regime. From comparison with theoretical predictions, CLG found that the dynamical 
mass of \ABDorC\, is almost twice than predicted by evolutionary models 
(Chabrier et al. 2000), which suggests that models tend to underpredict the mass of 
young VLM objects. In this context, a precise estimate of the dynamical mass of 
\ABDorC\, is extremely important. In this paper we report the details of 
an improved method to determine the mass of \ABDorC, which confirms the value 
of 0.090\,M$_\odot$ given by CLG. We also report on the sky motion 
of \ABDorBa, which 
shows a nearly-identical parallax to that of \ABDorA\, and evidence of the 
long-term orbital motion around \ABDorA. 

\section{Astrometric Data}

In Table 1 we summarize the available astrometric data of the 
AB\,Doradus system, which 
include absolute positions of \ABDorA\, and \ABDorBa, relative positions 
of the 9" pair \ABDorA\,/\ABDorBb, and relative positions of the closer 
pairs \ABDorA\,/\ABDorC\, and 
\ABDorBa\,/\ABDorBb. New absolute positions of \ABDorBa\, are presented in 
this table; they have been obtained from the same VLBI observations that 
were used to make the astrometric analysis of \ABDorA\, reported by 
Guirado et al. (1997). Given the 9" separation, \ABDorA\, and \ABDorBa\, 
lie within the primary beam of each of the telescopes 
and thus can be observed simultaneously for efficient cancellation of
atmospheric systematic errors. The interferometric array has 
much finer resolution (a few milliarcseconds) and, therefore, 
the interferometric data for \ABDorBa\, could be extracted and processed 
following the same procedures as 
described in Sect. 2 of Guirado et al. (1997) for \ABDorA. This in-beam 
technique is widely used in VLBI observations (e.g. Marcaide \& Shapiro 1983; 
Fomalont et al. 1999). On the other hand, the relatively low brightness 
of \ABDorBa\, ($V$=12.6; Collier Cameron \& Foing 1997) explains the absence of 
Hipparcos data for this star. In Sect. 3, we revisit the astrometry of the 
different pairs shown in Table 1.

\begin{table*}
\begin{minipage}{14cm}
\caption{Compilation of all available astrometric data for the AB\,Doradus 
system}
\begin{tabular}{lcccc}
\hline
\multicolumn{5}{c}{\ABDorA }\\
Epoch & Instrument & $\alpha$(J2000) & $\delta$(J2000) & Reference \\
\hline
1990.3888 & Hipparcos & $5^{h}\,28^{m}\,44\rlap{.}^{s}77474\,\pm\,0\rlap{.}^{s}00026$ &
                       $-65^{\circ}\,26'\,56\rlap{.}''2416\,\pm\,0\rlap{.}''0007$ & (1) \\
1990.5640 & Hipparcos & $5^{h}\,28^{m}\,44\rlap{.}^{s}78652\,\pm\,0\rlap{.}^{s}00025$ &
                       $-65^{\circ}\,26'\,56\rlap{.}''2272\,\pm\,0\rlap{.}''0007$ & (1) \\
1991.0490 & Hipparcos & $5^{h}\,28^{m}\,44\rlap{.}^{s}77578\,\pm\,0\rlap{.}^{s}00024$ &
                       $-65^{\circ}\,26'\,56\rlap{.}''2615\,\pm\,0\rlap{.}''0007$ & (1) \\
1991.5330 & Hipparcos & $5^{h}\,28^{m}\,44\rlap{.}^{s}78942\,\pm\,0\rlap{.}^{s}00025$ &
                       $-65^{\circ}\,26'\,56\rlap{.}''0757\,\pm\,0\rlap{.}''0008$ & (1) \\
1992.0180 & Hipparcos & $5^{h}\,28^{m}\,44\rlap{.}^{s}78202\,\pm\,0\rlap{.}^{s}00024$ &
                       $-65^{\circ}\,26'\,56\rlap{.}''1160\,\pm\,0\rlap{.}''0009$ & (1) \\
1992.2329 & VLBI      & $5^{h}\,28^{m}\,44\rlap{.}^{s}77687\,\pm\,0\rlap{.}^{s}00019$ &
                       $-65^{\circ}\,26'\,56\rlap{.}''0049\,\pm\,0\rlap{.}''0007$ & (1) \\
1992.6849 & VLBI      & $5^{h}\,28^{m}\,44\rlap{.}^{s}80124\,\pm\,0\rlap{.}^{s}00018$ &
                       $-65^{\circ}\,26'\,55\rlap{.}''9395\,\pm\,0\rlap{.}''0006$ & (1) \\
1993.1233 & VLBI      & $5^{h}\,28^{m}\,44\rlap{.}^{s}78492\,\pm\,0\rlap{.}^{s}00024$ &
                       $-65^{\circ}\,26'\,55\rlap{.}''9137\,\pm\,0\rlap{.}''0008$ & (1) \\
1994.8137 & VLBI      & $5^{h}\,28^{m}\,44\rlap{.}^{s}81768\,\pm\,0\rlap{.}^{s}00019$ &
                       $-65^{\circ}\,26'\,55\rlap{.}''6866\,\pm\,0\rlap{.}''0005$ & (1) \\
1995.1425 & VLBI      & $5^{h}\,28^{m}\,44\rlap{.}^{s}80247\,\pm\,0\rlap{.}^{s}00027$ &
                       $-65^{\circ}\,26'\,55\rlap{.}''6248\,\pm\,0\rlap{.}''0011$ & (1) \\
1996.1507 & VLBI      & $5^{h}\,28^{m}\,44\rlap{.}^{s}81137\,\pm\,0\rlap{.}^{s}00013$ &
                       $-65^{\circ}\,26'\,55\rlap{.}''4852\,\pm\,0\rlap{.}''0003$ & (1) \\
1996.3607 & VLBI      & $5^{h}\,28^{m}\,44\rlap{.}^{s}81776\,\pm\,0\rlap{.}^{s}00018$ &
                       $-65^{\circ}\,26'\,55\rlap{.}''3785\,\pm\,0\rlap{.}''0010$ & (1) \\
\hline
\multicolumn{5}{c}{\ABDorBa\, (=\Rst137B) }\\
Epoch & Instrument & $\alpha$(J2000) & $\delta$(J2000) & Reference \\
\hline
1992.2329 & VLBI      & $5^{h}\,28^{m}\,44\rlap{.}^{s}39520\,\pm\,0\rlap{.}^{s}0007$ &
                       $-65^{\circ}\,26'\,47\rlap{.}''0676\,\pm\,0\rlap{.}''0024$ & (2) \\
1992.6849 & VLBI      & $5^{h}\,28^{m}\,44\rlap{.}^{s}41973\,\pm\,0\rlap{.}^{s}0006$ &
                       $-65^{\circ}\,26'\,47\rlap{.}''0047\,\pm\,0\rlap{.}''0021$ & (2) \\
1993.1233 & VLBI      & $5^{h}\,28^{m}\,44\rlap{.}^{s}40441\,\pm\,0\rlap{.}^{s}0008$ &
                       $-65^{\circ}\,26'\,46\rlap{.}''9869\,\pm\,0\rlap{.}''0028$ & (2) \\
1994.8137 & VLBI      & $5^{h}\,28^{m}\,44\rlap{.}^{s}43687\,\pm\,0\rlap{.}^{s}0007$ &
                       $-65^{\circ}\,26'\,46\rlap{.}''5528\,\pm\,0\rlap{.}''0018$& (2)  \\
1996.1507 & VLBI      & $5^{h}\,28^{m}\,44\rlap{.}^{s}42842\,\pm\,0\rlap{.}^{s}0005$ &
                       $-65^{\circ}\,26'\,46\rlap{.}''5773\,\pm\,0\rlap{.}''0010$ & (2) \\ \hline
\multicolumn{5}{c}{Relative Position \ABDorA\, - \ABDorBa  }\\
Epoch & Instrument & Separation& P.A.\,($\degr$) & Reference \\
\hline
1929    & $-$    & $10\rlap{.}''0$ & $339$ & (3) \\ 
1985.7  & AAT    & $9\rlap{.}''3\,\pm\,0\rlap{.}''3$ & $344\,\pm\,5$ & (4) \\ 
1993.84 & ATCA   & $8\rlap{.}''90\,\pm\,0\rlap{.}''02$ & $345.2\,\pm\,0.1$ & (5) \\ 
1994.2  & Dutch/ESO   & $8\rlap{.}''9\,\pm\,0\rlap{.}''1$ & $344.7\,\pm\,0.3$ & (6) \\ 
2004.093 & VLT/NACO   & $9\rlap{.}''01\,\pm\,0\rlap{.}''01$ & $345.9\,\pm\,0.3$ & (7) \\ \hline

\multicolumn{5}{c}{Relative Position \ABDorA\, - \ABDorC  }\\
Epoch & Instrument & Separation& P.A.\,($\degr$) & Reference \\
\hline
2004.093 & VLT/NACO    & $0\rlap{.}''156\,\pm\,0\rlap{.}''010$ & $127\,\pm\,1\degr$ & (7) \\ \hline
\multicolumn{5}{c}{Relative Position \ABDorBa\, - \ABDorBb  }\\
Epoch & Instrument & Separation& P.A.\,($\degr$) & Reference \\
\hline
2004.098 & VLT/NACO    & $0\rlap{.}''062\,\pm\,0\rlap{.}''003$ & $236.4\,\pm\,3.33\degr$ & (8) \\ \hline

\end{tabular}

{\footnotesize (1) Guirado et al. (1997); (2) this paper; 
(3) Jeffers et al. (1963);
(4) Innis et al. (1986);
(5) J. Lim, personal communication 
(6) Mart\'{\i}n \& Brandner (1995); 
(7) Close et al. (2005);
(8) Brandner et al. in preparation } 

\end{minipage}
\end{table*}

\section{Astrometric Analysis}

\subsection{\ABDorA\,/\ABDorC\,: Orbit Determination}

The infrared image of \ABDorC\, provided the astrometric data that was
used by CLG to constrain the elements of the reflex orbit. The weakness 
of this procedure was that the relative position \ABDorA/\ABDorC\, was not 
included in the fit, rather it was only used as a discriminator of the orbits 
that plausibly fit the VLBI/Hipparcos data. In this section, 
we re-estimate the mass of \ABDorC\, using a much improved method that estimates 
the reflex orbit of \ABDorA\, by simultaneously combining both the existing 
VLBI/Hipparcos \ABDorA\, astrometric data and the 
near-infrared relative position of \ABDorC. 
Following the classical approach, we modeled the (absolute) position 
of \ABDorA\, ($\alpha$, $\delta$) at epoch $t$ from the 
expressions: 

\begin{eqnarray}
\lefteqn{ \alpha(t) = \alpha(t_{0}) + \mu_{\alpha}(t-t_0) + \pi\,P_{\alpha} } \nonumber\\ 
    & \qquad\qquad\;\; & +\:\orb_{\alpha}(t,X_1,X_2,X_3,X_4,P,e,T_0)  \nonumber \\
\lefteqn{ \delta(t) = \delta(t_{0}) + \mu_{\delta}(t-t_0) + \pi\,P_{\delta} \nonumber} \\ 
    & \qquad\qquad\;\; & +\:\orb_{\delta}(t,X_1,X_2,X_3,X_4,P,e,T_0) 
\end{eqnarray}

\noindent
where $t_0$ is the reference epoch, $\mu_{\alpha}$, $\mu_{\delta}$ are the
proper motions in each coordinate, $\pi$ is the parallax, $P_{\alpha}$ 
and $P_{\delta}$ are the parallax factors (e.g. Green 1985), and 
$\orb_{\alpha}$ and $\orb_{\delta}$ are the reflex orbital motions in 
$\alpha$ and $\delta$, respectively.  The astrometric parameters 
($\alpha(t_{0})$, $\delta(t_{0})$, $\mu_{\alpha}$, $\mu_{\delta}$, and
$\pi$) are linear in Eq. (1). 
The reflex motion, $\orb_{\alpha}$ and $\orb_{\delta}$, depends on the 
seven orbital parameters, namely, $a$, $i$, 
$\omega$, $\Omega$, $P$, $e$, and $T_0$. We have used the Thieles-Innes 
coefficients (Green 1985), represented by $X_1$, $X_2$, $X_3$, $X_4$, 
which are defined as combinations of $a$, $i$, $\omega$, $\Omega$. These coefficients
behave linearly in Eq. (1), leaving only three non-linear 
parameters ($P$, $e$, and $T_0$) to solve for in our weighted-least-squares 
approach.\\

\noindent
Since our fitting procedure estimates the orbital parameters of the 
reflex motion of \ABDorA, the 
relative separation \ABDorA/\ABDorC\, provided by the infrared 
data ($\Delta\alpha'$, $\Delta\delta'$) at epoch $t'$ is 
included in the fit via the corresponding orbital position of 
the primary star according to the definition of the center of mass 
of the system:

\begin{eqnarray}
\qquad\qquad\qquad\Delta\alpha'& = & -(1+q^{-1})\orb_{\alpha}(t') \nonumber \\
\Delta\delta'& = & -(1+q^{-1})\orb_{\delta}(t')  
\end{eqnarray}

\noindent
where $q$ is the mass ratio $m_c/m_a$, with $m_a$ being the mass of the primary 
and $m_c$ the mass of the companion. The combination of data types  
in the same fit is reflected in the definition of 
the ${\chi}^{2}$ to be minimized: 

\begin{eqnarray}
\lefteqn{ \chi^{2} = 
\sum_{i=1}^{N}
\frac{(\alpha(t_i)-\widehat{\alpha}(t_i))^2}{\sigma_{\alpha}^2(t_i)} \, + \,  
\sum_{i=1}^{N}
\frac{(\delta(t_i)-\widehat{\delta}(t_i))^2}{\sigma_{\delta}^2(t_i)} } \nonumber \\  
& &  \!\!+ \, (1+q^{-1})^2\bigg[\frac{(S_{\alpha}(t')-\widehat{S}_{\alpha}(t'))^2}
{\sigma_{S_{\alpha}}^2(t')}  + 
\frac{(S_{\delta}(t)-\widehat{S}_{\delta}(t'))^2}
{\sigma_{S_{\delta}}^2(t')}\bigg]
\end{eqnarray}

\noindent
where the $\sigma$'s are the corresponding standard deviations (Table 1) 
and the circumflexed quantities are the theoretical values of our  
{\it a priori} model. The virtue of the definition of $\chi^2$ in Eq. (3) is 
that the linearity of 
the orbital parameters is conserved as long as the mass ratio $q$ is 
not adjusted in the fit. In consequence, $m_a$ 
is a fixed parameter in our fit (we adopted the value of 
0.865$\pm$0.034\,M$_\odot$, as given by CLG). The mass of 
the secondary 
($m_c$) will be estimated via an iterative procedure that we outline 
below: 

\begin{enumerate}
\item We set {\it a priori} values of the three non-linear parameters 
($P$, $e$, and $T_0$). In particular, we sample the 
following parameter space: 
0$\,<\,P\,<$\,30\,\,years, 

1990.0\,$<\,T_0\,<$\,1990.0$\,+\,P$, and  
0$\,<\,e\,<\,$1. To start this procedure, we need an initial value of $m_c$. 
We take a value of 0.095\,M$_\odot$, which corresponds to 
the central value of the $m_c$ interval given in Guirado et al. (1997).

\item To find a minimum of $\chi^2$, as defined in Eq. (3), 
we used an iterative method, based on the Brent 
algorithm (Press et al. 1992).  
The minimum is defined such that the difference 
between the reduced-$\chi^2$ of two successive iterations is 
significantly less than unity. 

\item From the resulting orbital parameters, we then use 
Kepler's third law [$m^3_c/(m_a + m_c)^2=a_1^3/P^2$, with $a_1$ the 
semimajor axis of the reflex orbit] to 
estimate the mass $m_c$ of \ABDorC.

\item We iterate the least squares fit (step 2) using 
as {\it a priori} values the new set of adjusted orbital parameters, and 
estimated $m_c$. 

\item A final set of orbital parameters is obtained once 
the estimated $m_c$ is {\it self-consistent}, that is, 
the difference between the value of $m_c$ calculated in step 3 
from consecutive sets of adjusted orbital parameters is 
negligible (i.e. $\ll$0.001\,$M_{\odot}$). 

\end{enumerate}

\noindent
The resulting orbital parameters, and the estimate of the mass of \ABDorC, 
are shown in Table 2 and represented in Fig. 1. These values are fully compatible with those 
given by CLG. However, our method shows the robustness of the 
determined orbit. Despite the wide range of parameter space 
investigated, the solution found for the reflex orbit of \ABDorA\, is 
unique (see Fig. 2). This is a remarkable result: for astrometrically 
determined orbits, Black \& Scargle (1982) predicted a coupling 
between the proper motion and the orbital wobble, resulting in 
an underestimation of the period and semimajor axis. This coupling 
is present in the VLBI/Hipparcos data that only covers 51\% of the 
reflex orbit. Our least-squares approach copes partially with this effect, since 
the astrometric and orbital 
parameters are estimated {\it simultaneously}. However, the VLT/NACO data 
not only extends significantly the observing time baseline, but 
represents {\it purely} orbital information, independent of proper motion 
and parallax effects. In practice, the combination of astrometric data 
of the primary star with astrometric data of the relative orbit improves 
the fit dramatically, constraining the orbital periods allowed by the 
astrometric data of the primary only. In our case, the constraint is 
such that the only allowed period is 11.76$\pm$0.15\,yr. In general, 
our results show the combination of different 
techniques is more effective than any one technique alone.

\begin{figure}
\centering
\includegraphics[width=7cm]{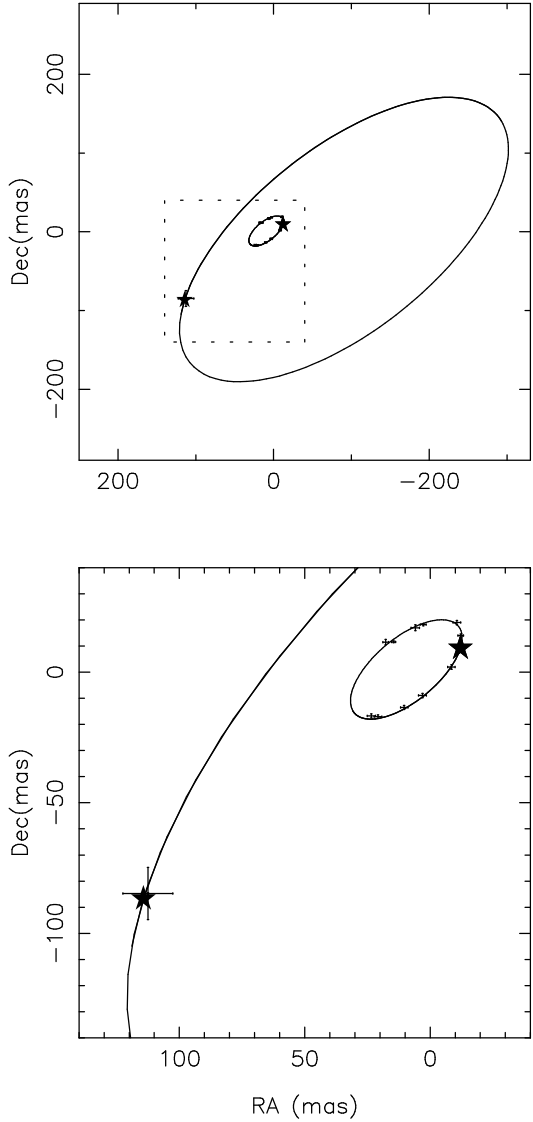}
\caption{Above: orbits of the pair \ABDorA\, (inner ellipse) and \ABDorC\, (outer ellipse). Below: blow up of the dotted square in the figure above. VLBI and 
Hipparcos data points are marked in AB\,Dor\,A's orbit, while the VLT/NACO 
\ABDorC\, position relative to \ABDorA\, is indicated in AB\,Dor\,C's ellipse. The 
star symbols over the 
orbits correspond the 
astrometric predictions at epoch 2004.093, based on the orbital elements given 
in Table 2.}
\end{figure}

\begin{table}
\begin{minipage}[t]{\columnwidth}
\caption{J2000.0 astrometric and orbital parameters of \ABDorA} 
\centering
\renewcommand{\footnoterule}{}
\begin{tabular}{ll}
\hline \hline
Parameter &   \\
\hline
$\alpha$
\footnote{The reference epoch is 1993.0. Units of right ascension are 
hours, minutes, and seconds, and units of declination are degrees, 
arcminutes, and arcseconds.}: 
& $5\,28\,44.7948$   \\
$\delta^a$: & $-65\,26\,55.933 $ \\
$\mu_{\alpha}$\,(s\,yr$^{-1}$): & $0.0077\pm 0.0002$  \\
$\mu_{\delta}$\,(arcsec\,yr$^{-1}$): & $0.1405\pm 0.0008$  \\
$\pi$\,(arcsec):          & $0.0664\pm  0.0005$ \\
   &\\
$P$\,(yr):           & $11.76\pm 0.15 $   \\
$a_1$\,(arcsec):     & $0.0322\pm  0.0002$  \\
$e$:                 & $0.60\pm 0.04 $  \\
$i$\,(deg):          & $67\pm 4 $  \\
$\omega$\,(deg):     & $109\pm 9 $ \\
$\Omega$\,(deg):     & $133\pm 2 $ \\
$T_o$:               & $1991.90\pm 0.04 $ \\
 &  \\
m$_{c}$\,(M$_\odot$)\footnote{
Mass range obtained from the period and semimajor axis 
via Kepler's third law. The mass adopted for the 
central star \ABDorA\, was 0.865$\pm$0.034\,M$_\odot$.}:
& $0.090\pm 0.003$ \\
\hline
\end{tabular}
\end{minipage}
\end{table}

\begin{figure}
\centering
\includegraphics[width=7cm]{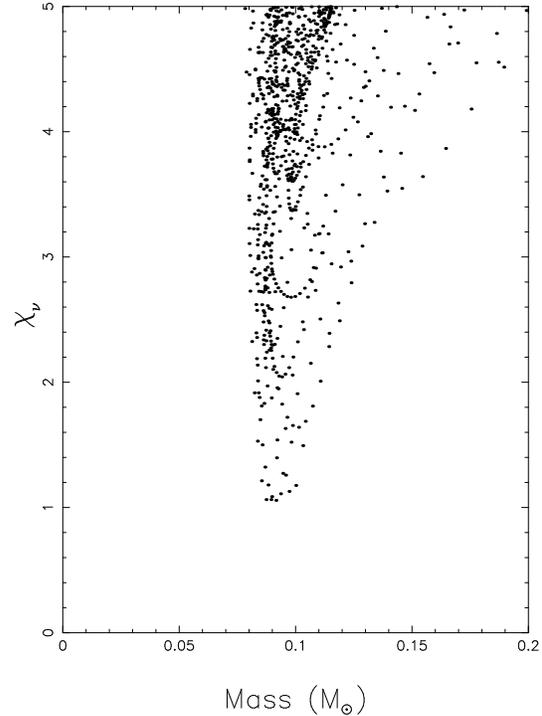}
\caption{Result of the exploration of the \ABDorA\, reflex orbit. A well-defined 
minimum is found for a mass companion of 0.090\,M$_{\odot}$. See Sect. 3.1.}
\end{figure}

\subsubsection{Error Analysis}

\noindent
Our least-squares procedure provides formal errors for the 
adjusted orbital parameters. However, other systematic 
contributions need to be taken into account. In particular, 
this includes the uncertainty associated to the mass of \ABDorA, which 
is a fixed parameter in our analysis 
(0.865$\pm$0.034\,M$_{\odot}$). To estimate this error contribution, 
we altered the mass of \ABDorA\, by one standard deviation 
and repeated the fitting procedure to obtain the change in the 
orbital parameters and the mass of \ABDorC. We note that this a 
conservative approach, since this technique 
fails to account for the correlation of $m_a$ with the rest 
of the parameters. The resulting parameter changes were added in quadrature 
with the 
formal errors of our fit (see Table 2). As expected, the 
0.003\,M$_{\odot}$ standard deviation of the mass 
of \ABDorC\, is dominated by the uncertainty in the mass 
of \ABDorA, while the standard deviations of the rest 
of the parameters are dominated by the statistical errors.\\
We also checked the dependence of our results on  
the choice of the {\it a priori} value of $m_c$ in step 1 of our 
fitting procedure (Sect. 3.1). We found that the results are 
insensitive to this choice. The postfit residuals of the positions 
of \ABDorA\, exhibits an rms of $\sim$1\,mas at each coordinate, 
consistent with the standard errors, and with no evidence, within 
uncertainties, of any further orbiting companion to \ABDorA.

\subsection{VLBI Astrometric Parameters of \ABDorBa}

\noindent
Innis et al. (1985) presented radial velocity measurements 
of \ABDorBa, the 9" companion to \ABDorA. Their measurements do not
differ from those of \ABDorA\, within the uncertainties. Additionally, 
Innis et al (1986) and 
Mart\'{\i}n \& Brandner (1995) reported close agreement between the proper 
motions of both stars. These results are strong arguments in favor of a
physical association of both stars. We used the VLBI (absolute) positions of 
\ABDorBa\, given in Table 1 to derive
the parallax and proper motion via a least-squares fit. 
The results of this fit are presented in Table 3, which shows that the parallax 
of \ABDorBa\, is coincident with that of \ABDorA\, to within
the uncertainties, which provides independent and conclusive evidence for the 
association of both stars. Comparison of Table 1 and Table 3 shows that 
the proper motion of \ABDorBa\, derived from the radio data appears 
significantly different to that of \ABDorA. Given the relatively small 
uncertainty of our determination, this does not contradict previous (and 
coarser) measurements of common proper motion. Rather, we interpret this 
extra proper motion of \ABDorBa\, towards the south-east as a result 
of the orbital motion around \ABDorA\, (see Sect. 3.3).

\begin{table}
\begin{minipage}[t]{\columnwidth}
\caption{J2000.0 VLBI astrometric parameters of \ABDorBa}
\centering
\renewcommand{\footnoterule}{}
\begin{tabular}{ll}
\hline \hline
Parameter &   \\
\hline
$\alpha$
\footnote{The reference epoch is 1993.0. Units of right ascension are 
hours, minutes, and seconds, and units of declination are degrees, 
arcminutes, and arcseconds.}: & $5\,28\,44.4123\pm 0.0002$   \\
$\delta\,\,^b$: & $-65\,26\,46.9974\pm 0.0015 $ \\
$\mu_{\alpha}$\,(s\,yr$^{-1}$): & $0.0085\pm 0.0002$  \\
$\mu_{\delta}$\,(arcsec\,yr$^{-1}$): & $0.134\pm 0.0012$  \\
$\pi$\,(arcsec):          & $0.0666\pm  0.0015$ \\
\hline
\end{tabular}
\end{minipage}

\end{table}

\noindent
The postfit residuals of \ABDorBa\, show a systematic signature,
both in right ascension and declination, which corresponds to a relatively 
high rms of $\sim$4\,mas. The short time span between our separate VLBI observations 
makes it unlikely that this signature is an effect of the long-term 
gravitational interaction of \ABDorBa\, with \ABDorA. Rather, this signature 
could be assigned to the 0.070" companion (ABDorBb) of \ABDorBa\, seen 
in the VLT/NACO observations reported by CLG.
As for the revision of the reflex orbit of \ABDorA,  
we attempted to get estimates of the orbital elements of the reflex 
motion of \ABDorBa\, by combining the radio data with 
the VLT relative position between 
\ABDorBa\,/\ABDorBb\, (Table 1). 
However, our analysis did not yield useful bounds to the 
mass of this pair, showing that the number of data points is 
still insufficient and, more likely, they do not properly sample 
the expected short period of this tight pair. 

\subsection{\ABDorA\,/\ABDorBa\,: evidence of orbital motion of \ABDorBa}

\noindent
As stated in the previous section, evidence of the motion of \ABDorBa\,
around \ABDorA\, can be obtained from the radio data alone. In order 
to get more precise evidence of this orbital motion, we 
augmented our data set with relative positions \ABDorA/\ABDorBa\, found in 
the literature (see Table 1). We then corrected all relative positions 
\ABDorA/\ABDorBa\, for the reflex orbital motion of \ABDorA\, (Table 2),
effectively referring the positions of \ABDorBa\, to the center of mass of 
the \ABDorA/\ABDorC\, system.\\
We attempted to constrain the relative orbit of \ABDorA\,/\ABDorBa\, following 
a similar 
analysis to that described in Sect. 3.1, fitting only the 7 parameters 
of the relative orbit. We sampled all possible periods up to 5000 years 
and eccentricities from 0 to 1. We selected as plausible orbits those 
whose reduced-$\chi^2$ differs by 25\% of the minimum. For each 
plausible orbit, the mass of the complete system was estimated
from Kepler's third 
law, now expressed in terms of the parameters of the relative orbit: \\

\begin{equation}
\qquad\qquad\qquad\frac{(a/\pi)^3}{P^2} = M_{(A+C)}+M_{(Ba+Bb)} 
\end{equation}

\noindent
where $M_{(A+C)}$ and $M_{(Ba+Bb)}$ are the combined masses of 
\ABDorA/\ABDorC\, 
and \ABDorBa\,/\ABDorBb, respectively, $a$ is the relative semimajor axis (arcsec), 
$\pi$ is the parallax (arcsec; Table 2), and $P$ is the period (yr). 
The poor coverage of the orbit favors a correlation between the 
orientation angles and the eccentricity, allowing a wide range of 
orbital parameters that fit our data equally well. However, a similar 
correlation between $P$ and $a$ imposes a constraint on the 
determination of the mass of the system via Eq. (4), which is 
represented in the histogram of Fig. 3. From the plausible orbits 
selected, more than 50\% correspond to a total mass of 
the AB\,Doradus system 
in the interval 0.95$-$1.35\,M$_\odot$ (see Fig. 4 for examples of 
plausible orbits). 
Larger masses are not excluded, 
but the required orbital configurations for masses outside this range 
occur with significantly reduced probability.\\
If we assume the total mass of the AB\,Doradus system lies in the interval 
0.95$-$1.35\,M$_\odot$, the combination with our estimate of 
$M_{(A+C)}$ (0.956$\pm$0.035\,M$_\odot$; see Sect. 1) suggests an upper 
bound to the mass of the pair \ABDorBa\,/\ABDorBb\, 
of 0.4\,M$_\odot$. This upper limit to $M_{(Ba+Bb)}$ looks too coarse 
to calibrate evolutionary models. Nevertheless, it can be transformed 
into a bound to the age of this pair. 
To do this, we used the $K$-band 2MASS photometry
of \ABDorBa, and the $K$-band difference between \ABDorBa\, and \ABDorBb\, 
reported by CLG. The comparison with 
Baraffe et al. (1998) isochrones suggests an  
age for this pair in the range of 50$-$120\,Myr. 
This range is compatible with 
previous values of the age of \ABDorBa\,
(30$-$100\,Myr; Collier Cameron \& Foing 1997).
However, our age estimate for \ABDorBa\,/\ABDorBb\,  
is not conclusive: 
first, the masses of the individual components are yet to 
be determined, and 
second, there are indications that the evolutionary models might need 
revision, since they tend to underpredict masses for very young objects below 
0.3\,M$_\odot$ (CLG; Reiners et al. 2005).

\begin{figure}
\centering
\includegraphics[width=7cm]{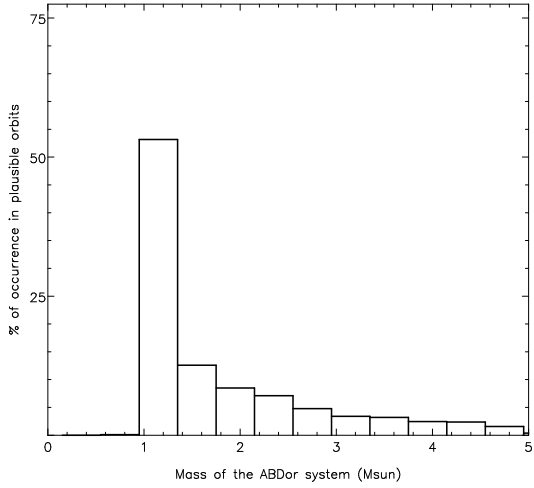}
\caption{Histogram of plausible orbits for the relative orbit of 
\ABDorBa\, around \ABDorA. More than 50\% of the plausible orbits 
correspond to a total mass of the system in the range 0.95$-$1.35\,M$_{\odot}$.}
\end{figure}

\begin{figure}
\centering
\includegraphics[height=14cm]{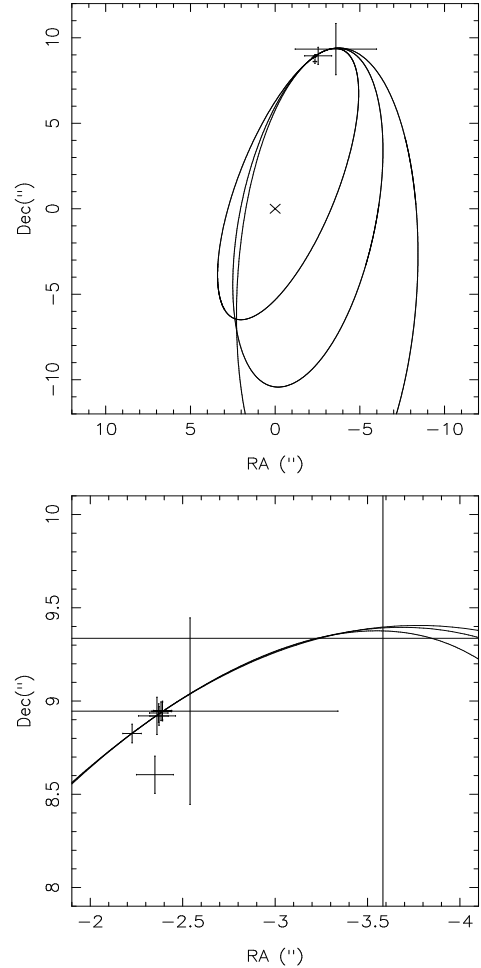}
\caption{Above: positions of \ABDorBa\, with respect to the center of mass of 
\ABDorA/\ABDorC\, (see Table 1) and several 
allowed orbital solutions. The displayed 
orbits correspond to a total mass 
of the system in the range 0.95-1.35\,M$_{\odot}$ with periods 
of 1400, 2300, and 4300 years. The cross at the origin indicates the 
position of \ABDorA/\ABDorC. Below: blow up of the region 
containing the measurements.}  
\end{figure}

\section{Summary}

We have revisited the different orbits in the quadruple system in 
AB\,Doradus. Paradoxically, this system, where 
the measurement of precise radial velocities is difficult due to 
the fast rotation of the main components, has
become an extraordinary target for astrometric techniques in 
different bands of the electromagnetic spectrum. 
From our analysis of the available data, we have re-estimated the mass of the 
VLM star \ABDorC\, by using a least-square approach that combines the data from 
radio, optical, and infrared bands. Although the data do not cover 
a full orbit, the mass and orbital elements of \ABDorC\, are 
strongly constrained and fully compatible with those reported by 
CLG. Further monitoring of the 
reflex orbit of \ABDorA\, via VLBI observations, and of the relative orbit 
\ABDorA\,/\ABDorC\, via VLT/NACO observations, will result in independent estimates 
of the masses of the components of this pair. 
From the absolute radio positions of \ABDorBa, we have 
determined the absolute sky motion (i.e. not referred to the motion of \ABDorA) 
of this star and, in particular, its parallax, which 
is identical, within the uncertainties, to that of \ABDorA. This confirms 
the association of both stars. 
The mass of \ABDorC\, serves as a precise calibration point for 
mass-luminosity relations of young VLM stars. Likewise, other components 
of AB\,Doradus may provide new calibration points for slightly higher masses. 
We have found evidence for the long-term orbital motion of 
\ABDorBa\,/\ABDorBb\, around \ABDorA/\ABDorC. From an exploration of the multiple 
orbits that fit the available data
we find that the most probable mass upper limit of the pair is 0.4\,M$_\odot$.
This limit maps into an age range of 50$-$120\,Myr using the isochrones provided 
by Baraffe et al. (1998). 
Further monitoring with the appropriate sampling, both 
in radio 
and infrared, should provide the orbital elements of both the relative 
and reflex orbits of the pairs \ABDorA\,/\ABDorC\, and \ABDorBa\,/\ABDorBb, 
from which would follow precise, model-independent, estimates of the masses 
of the four components of this system.


\begin{acknowledgements}
This work has been supported by
the Spanish DGICYT grant AYA2002-00897. 
The Australia Telescope is funded by the Commonwealth Government for 
the operation as a national facility by the CSIRO. 
Part of this research was carried 
out at the Jet Propulsion Laboratory, California Institute of 
Technology, under contract with the US National Aeronautics and 
Space Administration.
\end{acknowledgements}


\begin{thebibliography}{}

\bibitem[]{}
Baraffe, I., Chabrier, G., Allard, F., \& Hauschildt, P.H. 1998, ApJ, 499, 205 

\bibitem[]{}
Black, D.C., \& Scargle, J.D. 1982, ApJ, 263, 854 

\bibitem[]{}
Bouy, H., Duchene, G., K\"ohler, R., et al. 2004, A\&A, 423, 341

\bibitem[]{}
Chabrier, G., Baraffe, I., Allard, F., \& Hauschildt, P.H. 2000, ApJ, 542, 464 

\bibitem[]{}
Close, L.M., Lenzen, R., Guirado, J.C., et al. 2005, Nature, 433, 286 (CLG)

\bibitem[]{}
Collier Cameron, A., \& Foing, B. 1997, The Observatory, 117, 218

\bibitem[]{}
Fomalont, E.B., Goss, W.M., Beasley, A.J. \& Chatterje, S. 1999, AJ, 117, 3025

\bibitem[]{}
Golimowski, D.A., Henry, T.J., Krist, J.E., et al. 2004, ApJ, 128, 1733

\bibitem[]{}
Green, R.M. 1985, Spherical Astronomy (Cambridge: Cambridge University Press) 

\bibitem[]{}
G\"udel, M., Audard, M., Briggs, K., et al. 2001, A\&A, 365, L336

\bibitem[]{}
Guirado, J.C., Reynolds, J.E., Lestrade, J.-F., et al. 1997, ApJ, 490, 835 

\bibitem[]{}
Innis, J.L., Coates, D.W., Thompson, K. \& Robinson, R.D. 1985, PASAu, 6, 156 

\bibitem[]{}
Innis, J.L., Thompson, K. \& Coates, D.W. 1986, MNRAS, 223, 183 

\bibitem[]{}
Jeffers, H.M., van der Bos, W.H., \& Greeby, F.M. 1963, Index Catalogue of Visual
Double Stars, Lick Observatory, Mount Hamilton, California 

\bibitem[]{}
Kovalevsky, J. 1995, Modern Astrometry (Berlin: Springer) 

\bibitem[]{}
Lane, B.F., Zapatero Osorio, M.R., Britton, M.C., et al. 2001, ApJ, 560, 390 

\bibitem[]{}
Lim, J., Nelson, G.J., Castro, C., et al. 1992, ApJ, 405, L33 

\bibitem[]{}
Lim, J. 1993, ApJ, 405, L33 

\bibitem[]{}
Mart\'{\i}n, E., \& Brandner, W. 1995, 294, 744 

\bibitem[]{}
Marcaide, J.M. \& Shapiro, I.I. 1983, 88, 1133 

\bibitem[]{}
Mewe, R., Kaastra, J.S., White, S.M. \& Pallavicini, R. 1996, A\&A, 315, 170

\bibitem[]{}
Press, W.H., Teukolsky, S.A., Vetterling, V.T., \& Flannery, B.P. 1992, 
Numerical Recipes in FORTRAN: The Art of Scientific Computing 
(Cambridge: Cambridge Univ. Press) 

\bibitem[]{}
Reiners, A., Basri, G. \& Mohanty, S. 2005, ApJ, in press (arXiv:astro-h/0506501)

\bibitem[]{}
Vilhu, O., Muhli, P., Huovelin, J. et al. 1998, ApJ, 115, 1610

\end{thebibliography}
\end{document}